\documentclass[10pt]{article}

\usepackage{amsmath}
\usepackage{amssymb}
\usepackage{graphicx}
\usepackage{cite}
\usepackage[usenames]{color}
\usepackage{units}

\usepackage[labelfont=bf,labelsep=period,justification=raggedright]{caption}

\makeatletter
\renewcommand{\@biblabel}[1]{\quad#1.}
\makeatother

\date{}

\ifdefined\debug
  \usepackage{ulem} 
  \newcommand{\added}[1]{{\bf\color{OliveGreen}#1}}
  
  \newcommand{\changed}[1]{{\bf\color{Blue}#1}}
  \newcommand{\deleted}[1]{{\bf\color{Red}\sout{#1}}}
  
\else
  \newcommand{\added}[1]{#1}
  
  \newcommand{\changed}[1]{#1}
  \newcommand{\deleted}[1]{}
\fi

\newcommand{\comment}[1]{}

\def\showimages{}
\ifdefined\showimages
  \newcommand{\showimg}[1]{#1}
\else
  \newcommand{\showimg}[1]{}
\fi

\begin{document}

\begin{flushleft}
  {\Large
    \textbf{Constructing a Stochastic Model of Bumblebee Flights from Experimental Data}
  } 
  \\
  Friedrich Lenz$^{1,\ast}$, 
  Aleksei V. Chechkin$^{2,3}$, 
  Rainer Klages$^{1}$
  \\
  \bf{1} School of Mathematical Sciences, Queen Mary University of London, Mile End Road, London E1 4NS, UK
  \\
\changed{  \bf{2} Max-Planck Institute for Physics of Complex Systems, Noethnitzer Str. 38, 01187 Dresden, Germany }
  \\
  \bf{3} Institute for Theoretical Physics, NSC KIPT, ul.\ Akademicheskaya 1,UA-61108 Kharkov, Ukraine
  \\
  $\ast$ E-mail: f.lenz@qmul.ac.uk
\end{flushleft}

\section*{Abstract}
  The movement of organisms is subject to a multitude of influences of widely varying character:
  from the bio-mechanics of the individual, over the interaction with the complex environment many animals live in,
  to evolutionary pressure and energy constraints.
  As the number of factors is large, it is very hard to build comprehensive movement models.
  Even when movement patterns in simple environments are analysed, the organisms can display very complex behaviours.
  While for largely undirected motion or long observation times the dynamics can sometimes
  be described by isotropic random walks, usually the directional persistence due to a preference
  to move forward has to be accounted for, e.g., by a correlated random walk.
  In this paper we generalise these descriptions to a model in terms of stochastic    
  differential equations of Langevin type, which we use to analyse experimental search flight data of foraging bumblebees.
  Using parameter estimates we discuss the differences and similarities to correlated random walks.
  From simulations we generate artificial bumblebee trajectories which we use
  as a validation by comparing the generated ones to the experimental data.


\section*{Introduction}
\subsection*{Foraging Animals}

  The characteristics of the movement of animals play a key role in a variety of ecologically relevant processes,
  from foraging and group behaviour of animals\cite{Santos2009} to dispersal\cite{Petrovskii2009,Hawkes2009}
  and territoriality\cite{Giuggioli2012}.
  Studying the behaviour of animals, simple random walk models have been proven effective in describing
  irregular paths\cite{Codling2008}.
  While the first studies on random paths of organisms focused on uncorrelated step sequences\cite{Pearson1905},
  in many cases of studies of animal behaviour the directional persistence of the animals suggested a modelling
  in terms of correlated random walks (CRWs) \cite{Kareiva1983,Bovet1988}.
  In many complex environments an intermittent behaviour of animals is observed.
  In these cases an animal switches either randomly or in reaction to its environment between different movement patterns.
  The mechanisms which generate, and the factors which influence this switching behaviour have been shown to be important
  in understanding and modelling complicated animal paths\cite{Benichou2006,Plank2009,Mashanova2010,Benichou2011review}.
  While there is a source of switching between free flight and food inspections in the experiment we analyse\cite{ourPRL},
  here we concentrate on the former as detailed below.
  With no clear indication of additional intermittency, we will focus on non-intermittent models in the following.

\subsection*{CRW/Reorientation Model}

  The planar horizontal movement of an animal is often approximated by a sequence of steps:
  an angle $\alpha(t)$ describes the current direction of movement in a fixed coordinate frame,
  while the step length $l(t)$ determines the distance travelled during a time step.
  The direction $\alpha(t)$, often determined by a specific
  front direction of the animal, changes each time step by a random turning angle $\beta(t)$.
  The description of the dynamics in a co-moving frame, i.e.,
  via the turning angle, turned out to be most useful for analysis of
  persistent animal movement \cite{Kareiva1983,Bovet1988}.
  In many cases $\beta(t)$ is drawn independently and identically distributed (i.i.d.)
  from a wrapped normal distribution or a von Mises distribution \cite{Codling2005,Batschelet1981} for
  each time step, giving rise to a persistence in direction depending on how strongly the distribution
  is concentrated around $0$.
  Usually the step length is taken to be either constant or it is drawn i.i.d.\ from some distribution.
  The step length can either be the result of a constant speed and a variable time step or
  (as in our case below) of a constant time step $\Delta t$ and a variable speed $s(t)$.

  This class of models can generate a variety of different dynamics. 
  Two special cases with a uniform distribution for $\beta$ and a fixed time step $\Delta t$
  are the standard Gaussian random walk for
  step lengths $l(t)=|z(t)|$ where $z$ is normally distributed
  and L\'evy flights for power-law tails in the step lengths distributions ($l(t)\sim l^{-\mu}$
  for $1<\mu\leq3$ and $l>l_0$).
  Related to L\'evy Flights, but using a time step proportional to the step length, are L\'evy Walks,
  which have been of interest as candidates for optimal search behaviour of foraging animals. They have been studied
  analytically\cite{ViswanathanStanley1999}, by simulations\cite{pitchford-efficient-or-inaccurate,Plank2009},
  and many experimental data sets have been statistically analysed to determine whether L\'evy Walks are suitable to describe
  the movement of animals (see, e.g., \cite{Viswanathan1996,Edwards2007,Benhamou2007,Sims2010,overturnFishingEdwards2011}).

  As L\'evy-type models show anomalous diffusive behaviour, in contrast to models with a finite variance of the step length
  distribution and a fixed time step $\Delta t$, only the latter are included in the definition of
  \emph{correlated random walks} which are also called \emph{reorientation models} in the context of animal movement. 
  Apart from pathological cases, CRWs are diffusive in the long time limit according to the central limit theorem.

  The estimation of the tortuosity of a trajectory is intimately connected to the distributions
  of the turning angle and speed \cite{Bovet1988,Codling2005,Benhamou2004}.
  The relevance of the turning angle distribution for foraging efficiencies when searching in random environments
  has been analysed, e.g., in \cite{Bartumeus2008}.

\subsection*{Generalisation of the Model}

  In the following we will present a generalisation of the CRW above, which we then use to analyse bumblebee flight data.
  Given movement data with a constant time step $\Delta t$, the step length is determined by the speed $s(t)=|v(t)|$ of the animal.
  As we will be looking at a flying insect in a data recording using a small time step,
  we may expect to have a deterministic persistence due to the animals momentum.
  Additionally, the above CRW model assumes that $s$ and $\beta$ are drawn i.i.d.\ which is sensible
  if $\Delta t$ is large enough.
  However, for small time steps it cannot be excluded that the decision of the animal to turn left or right takes longer
  than the time step, which can correlate the turning angles $\beta(t)$ over a number of time steps.
  To allow for these possibilities we therefore model the changes in speed and turning angle via two coupled
  generalized Langevin equations,
    \begin{align}
      \frac{d\beta}{dt}(t) &= h(\beta(t),s(t)) + \tilde{\xi}_s(t) \label{eq:generalmodel-beta}\\
      \frac{ds}{dt}(t)&= g(\beta(t),s(t)) + \psi(t) ,           \label{eq:generalmodel-s}
  	\end{align}
  where we distinguish between the deterministic parts $h$ and $g$
  and stochastic terms $\psi$ and $\tilde{\xi}_s$ (whose speed dependency will be discussed in the Results section).
  We assume that the noise processes are stationary with auto-correlation functions which may be non-trivial, and we make no further assumptions for the shape of their stationary distributions.
  While Eqs.~(\ref{eq:generalmodel-beta},\ref{eq:generalmodel-s}) represent
  a time-continuous description, the turning angle $\beta$ still yields
  the change of $\alpha$ according to our fixed time resolution $\Delta t$.
  That is, $\beta(t)$ relates to a time-continuous angular velocity $\gamma$ of $\alpha$ via
  $\beta(t) = \int_{t-\Delta t}^t \gamma(\tau)d\tau$.  
  The animals' position $r(t)=(x(t),y(t))$ is then given by
  $dx/dt      = s \cos(\alpha(t))$,
  $dy/dt      = s \sin(\alpha(t))$ and
  $d\alpha/dt = \gamma(t)$.
  \added{Not having experimental access to $\gamma$,} the numerical analysis is done with time-discrete data where
  the measured turning angle is given by
  $\beta(t) =\added{\alpha(t)-\alpha(t-\Delta t) =} \measuredangle(v(t),v(t-\Delta t))$, where
  $v(t)=(r(t+\Delta t)-r(t))/\Delta t$
  at times $t=n\Delta t$, $n\in \mathbb{N}$.

\subsection*{Application to Experimental Data}
\begin{figure}[t] 
	\begin{center}
    \showimg{ \includegraphics[scale=0.5]{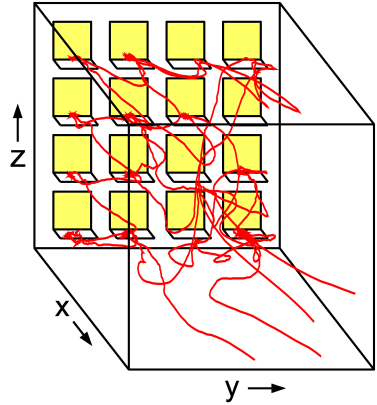} }
	\end{center}
	\caption{
      {\bf Sketch of the foraging arena together with part of the flight trajectory of
      a single bumblebee.}
      The bumblebees forage on a grid of artificial flowers on one wall of the
      box. While being on the landing platforms, the bumblebees have access to
      food supply. Of interest in this paper is the movement when the bumblebee is \emph{not} near the flower wall.
	}\label{fig:cage}
\end{figure}

  Analysing measured movement data of animals in their natural habitat is intricate due to
	a variety of factors which may influence the animal's behaviour,
	ranging from heterogeneous food source distributions \cite{Boyer2012,Kai2009,Sims2012}
	and predation threats\cite{Reynolds2010,ourPRL} to \changed{individual differences in behaviour within a population} \cite{Petrovskii2009,Hawkes2009}.
	Here we analyse data obtained from a small scale laboratory experiment in which single bumblebees forage in an artificial
	flight arena\cite{Ings2008}.
	The set-up is shown in Fig.~\ref{fig:cage} together with part of a typical trajectory of a bumblebee on its search for food.
  Each bumblebee can forage on an artificial flower carpet which is positioned on one of the walls of the arena.
	In this paper we are not interested in the behaviour resulting from the interaction with the flowers which has been studied
	in detail in \cite{ourPRL}. Instead we only examine the search flights away from the flower carpet.	(See section Materials and Methods for details.)
	We use our generalised stochastic model (Eqs.~(\ref{eq:generalmodel-beta},\ref{eq:generalmodel-s})) to describe these flights
	and to examine in which ways the behaviour deviates from a simple CRW model.
  Here we will focus on the horizontal movements. By neglecting the slower vertical movements, which are of more interest when
  analysing the starting and landing behaviour near flowers, we thus restrict ourselves to a two-dimensional model.

\section*{Results and Discussion}
\subsection*{Estimation of Drift Terms}
\begin{figure}[t] 
	\begin{center}
    \showimg{ \includegraphics[scale=0.6,angle=0]{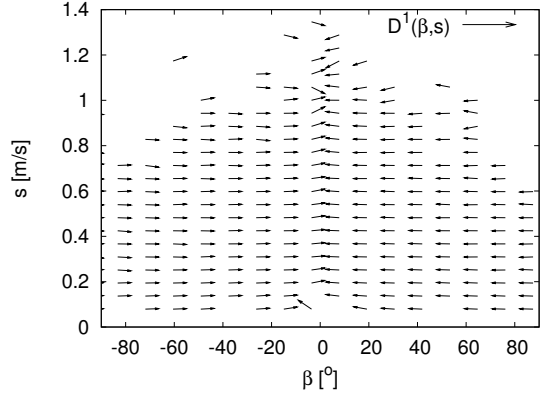} }
	\end{center}
	\caption{
	{\bf Normalised drift vector field $D^1(\beta,s)$} corresponding to the deterministic terms of the Langevin equations (Eqs.~(\ref{eq:generalmodel-beta},\ref{eq:generalmodel-s})) estimated via
	Eq.~(\ref{eq:drift}).
	The regular structure shows the quick relaxation to small angles\added{,} and the absence of strong cross-dependencies
	in the drift\added{, i.e., the $\beta$-dependence of the $s$-component of the vectors is weak and vice versa}.
	}\label{fig:fpc}
\end{figure}

  Given the experimental data, we start determining the unknown parameters in our model
  by first estimating the deterministic parts $h(\beta,s)$ and $g(\beta,s)$ of the Langevin equation.
  This is done by numerical estimation \cite{Risken,FriedrichPeinke1997,Ragwitz2001,Lenz2009} of the components
  of the drift vector field (drift coefficients) $D^1(\beta,s)=(g(\beta,s),h(\beta,s))^\top$
  of the corresponding Fokker-Planck equation via
  \begin{equation}\label{eq:drift}
    D^1(X) = \lim_{\tau\rightarrow 0}\frac{1}{\tau} \left.\left<\widetilde X(t+\tau)-X\right>\right|_{\widetilde{X}(t)\approx X}
  \end{equation}
  where $X=(\beta,s)^\top$ and $\left<.\right>$ is the time average over the time series $\widetilde X$
  conditioned on $\widetilde{X}(t)\approx X$, where $\widetilde{X}$ is assumed to be stationary
  (for a detailed discussion see \cite{Ragwitz2001}).
  The estimation of the drift terms is based on a Markov approximation:
    only those parts of the dynamics which match to a Markovian description in the state space variables $\beta$ and $s$ 
    have their deterministic terms reflected in $D^1(X)$.
    Any other parts of the flight dynamics -- stochastic as well as deterministic
    but not Markovian in $\beta$ and $s$ -- are captured by the stochastic terms
    of Eqs.~(\ref{eq:generalmodel-beta},\ref{eq:generalmodel-s}).
  Figure~\ref{fig:fpc} shows the drift vector field, with normalised lengths of the vectors for better visibility.
  The nearly horizontal vectors show, that the drift quickly pushes the turning angle $\beta$ towards $0$,
  while the dynamics in the speed $s$ is much slower.
  As the cross-dependencies of $h(\beta,s)$ on $s$ and of $g(\beta,s)$ on $\beta$ are weak,
  we can neglect them in our model.
  Since vector fields are hard to interpret, we will look at the projections in the following.

\begin{figure}[h] 
  \begin{center}
    \showimg{ \includegraphics[scale=0.6,angle=0]{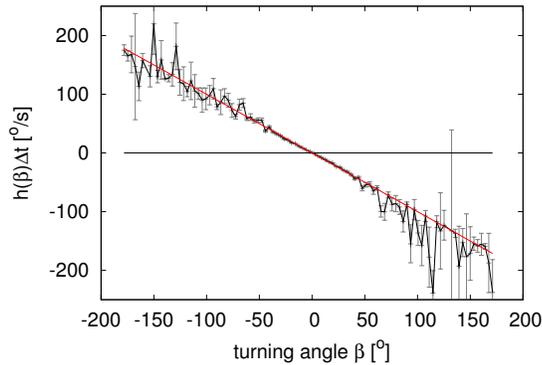} }
	\end{center}
	\caption{
	{\bf Drift coefficient of turning angle.}
	The deterministic drift $h(\beta)$ as estimated from data (black) is in good approximation (red) linear in $\beta$.
	(95\% confidence intervals in grey)
	}\label{fig:anglefpc}
\end{figure}

  Examining the drift $h(\beta)$ of the turning angle in Fig.~\ref{fig:anglefpc} reveals that the drift term
  seems linear in $\beta$ --- indeed we find numerically that its slope $-k$ matches exactly to a decay of the turning angle to $0$ in a single observation time step $\Delta t$ by $k \approx 1/\Delta t$, disregarding the noise term.
  This means that by integrating Eq.~(\ref{eq:generalmodel-beta}) over a time $\Delta t$
   and approximating the drift $h(\beta)$ for small $\Delta t$ by
   $ \int_t^{t+\Delta t} h(\beta(\tau)) d\tau \approx h(\beta(t))\Delta t $,
   we have
  \begin{equation}\label{eq:simpl}
   \beta(t+\Delta t)-\beta(t)= -k \beta(t)\Delta t + \int_t^{t+\Delta t}\tilde{\xi}_s(\tau)d\tau
                                 = -  \beta(t)         + \int_t^{t+\Delta t}\tilde{\xi}_s(\tau)d\tau \; .
  \end{equation} 
  With $ \xi_s(t) := \int_{t-\Delta t}^t \tilde{\xi}_s(\tau) d\tau $ and Eq.~(\ref{eq:simpl}),
  the time scale separation in the $\beta$-Langevin equation due to the very fast relaxation means
  that we can simplify Eqs.~(\ref{eq:generalmodel-beta},\ref{eq:generalmodel-s}) to:
  \begin{align}
    \beta(t) &= \xi_s(t)                   \label{eq:completemodel-beta}\\
    \frac{ds}{dt}(t)&= g(s(t)) + \psi(t) \label{eq:completemodel-s} .
  \end{align}
  \comment{While this reduction of dynamics from $d\beta/dt$ to $\beta$ goes into the direction of the simple
  reorientation model, the turning angles are still correlated, as we will see below.}
  \changed{While this reduction of the turning angle dynamics from $d\beta/dt$ to $\beta$
      bears similarity to a simple reorientation model,\comment{ we emphasize that}
      the turning angles are still correlated and speed-dependent, as we will see below.}

\begin{figure}[h]  
	\begin{center}
    \showimg{ \includegraphics[scale=0.6,angle=0]{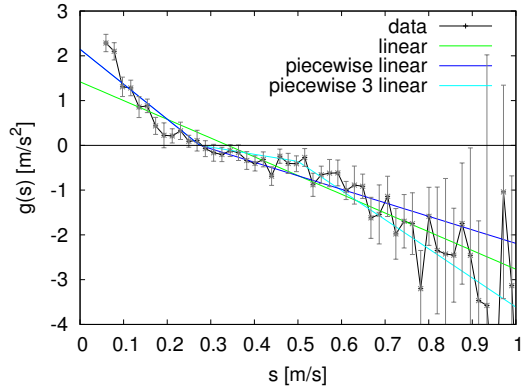} }
	\end{center}
	\caption{
	{\bf Drift coefficient of speed.}
	The experimental deterministic drift coefficient $g(s)$ (black, 95\% confidence intervals in grey) has been approximated by piecewise linear functions from one to three pieces (blue,green,cyan). The data shows the tendency to quickly increase low speeds. However, speeds above \unit[0.27]{m/s} decrease more slowly, except for the rare high speeds.
	}\label{fig:speedfpc}
\end{figure}

  The speed drift $g(s)$ displayed in Fig.~\ref{fig:speedfpc} shows that the deterministic part
  of the speed-Langevin equation alone would have a stable fixed point around $s_0=\unit[0.27]{m/s}$.
  Comparing the slopes above and below $s_0$ reveals that for $s<s_0$ the force towards $s_0$ is stronger than for $s>s_0$.
  This is biologically plausible if one interprets $s_0$ as a preferred speed: if the bumblebee is slower it accelerates,
  but if it is faster it does not rush to decelerate as it would give up the energy spent to reach a high velocity.
  For very high velocities (over \unit[0.55]{m/s}) the slope of $g(s)$ increases again.
  This might be caused by the limited space available to the bumblebee in the flight arena.
  For our model we approximated $g(s)$ by a piecewise linear function:
  \begin{equation}\label{eq:g-piecewise}
    g(s)\approx (s-s_0)\times	\left\{ \begin{array}{rcl}
                            						-d_1 & \mbox{for}& s<s_0 \\
    					                        	-d_2 & \mbox{for} & s\geq s_0
                          						\end{array}\right.  , 
  \end{equation}
  where $d_1>d_2>0$. As the very high velocities are rare, it made no difference in our model whether we used
  Eq.~(\ref{eq:g-piecewise}) or a piecewise linear function with three pieces.
						
\subsection*{Velocity-Dependent Angle-Noise and Noise Auto-Correlations}

\begin{figure}[h] 
	\begin{center}
    \showimg{ \includegraphics[scale=0.4]{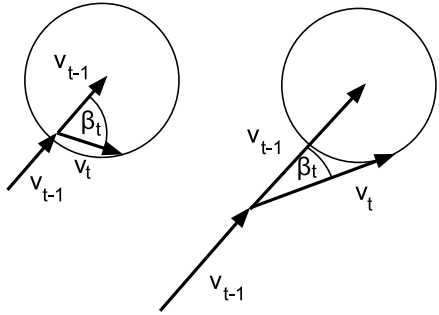}	
	  }
	\end{center}
	\caption{
	{\bf Schematics of the dependence of $\beta$ on speed $s$.}
	Assuming a constant maximal force (circle) available to the bumblebee to accelerate during a time step, the
	distribution of the turning angle $\beta$ depends on the previous speed $s_{t-1}=|v_{t-1}|$.
	Illustrated is the change from large angles for low speeds (left) to a stronger concentration around $0^\circ$	for higher speeds (right).
	}\label{fig:schematic}
\end{figure}

\begin{figure}[h] 
	\begin{center}
    \showimg{\includegraphics[scale=0.6,angle=0]{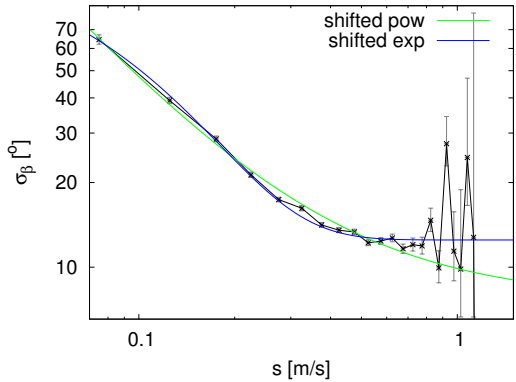} }
	\end{center}
	\caption{
	{\bf Log-log plot demonstrating the speed dependence of the turning angle distribution.}
	The standard-deviation $\sigma_\beta$ of the turning angle is shown as a function of the speed
	as estimated from data (black) and approximated by a shifted power-law (green) and a shifted exponential (blue).
  95\% confidence intervals for $\sigma_\beta$ based on a $\chi^2$-distribution are shown in grey.
	}\label{fig:beta-s-loglog}
\end{figure}

  What we did not specify before was that the turning angle distribution may depend on the speed of the bumblebees. 
  Given that the force a bumblebee can use to change directions is finite, the largest turning angles have to be smaller when
  flying with high speeds (see Fig.~\ref{fig:schematic}).
  This is consistent with the absence of simultaneously having high speed and large turning angle in the data - as is evident,
  e.g., from the data gaps in Fig.~\ref{fig:fpc}. However, animals can counteract this geometric dependence by varying the
  forces used for changing direction with the speed.
  We approximated the distribution for the turning angles for each speed $s$ by a normal distribution.
  This approximation works best for low speeds. While there are some deviations for high speeds,
  it was not possible to reliably fit a better model due to the limited amount of data available.
  Figure~\ref{fig:beta-s-loglog} shows how its standard deviation $\sigma_{\beta}$ depends on the current speed.
  $\sigma_{\beta}$ decreases with increasing speed, however it does not decay to $0$ as a simple geometric model would predict
  (see Materials and Methods below).

  Instead $\sigma_{\beta}(s)$ decays roughly exponentially to a constant offset.
  We therefore model the turning angles as speed-dependent
  \deleted{Gaussian} noise \added{with a wrapped normal distribution \cite{Codling2005,Batschelet1981}}:
  $\xi_s(t)\sim\mathcal{N}(0,\sigma_\xi(s))$ with $\sigma_\xi(s)= c_1 e^{-c_2 s}+c_3$.
  This offset could either be an effect of the boundedness of the flight arena,
  since the bumblebee has to turn more often to avoid walls when flying fast.
  Or it could be that the bumblebees use stronger forces for turning during fast flights to maintain their manoeuvrability.
  It would be interesting to examine free-flight data to check for the cause.
  In other models in which the momentum of the animal is not important for the observed directional persistence,
  this cross-dependence is often neglected \cite{Kareiva1983}.

\begin{figure}[h] 
	\begin{center}
 	  \showimg{ \includegraphics[scale=0.55,angle=0]{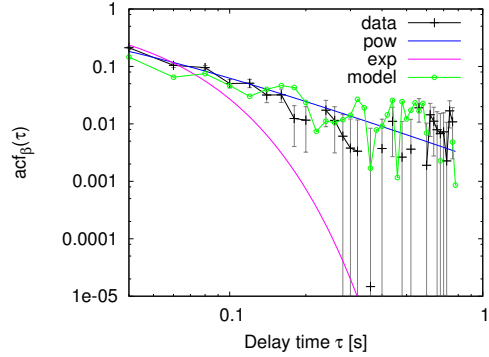} }
	\end{center}
	\caption{
	{\bf Log-log plot of auto-correlation of turning angles $\beta$.}
	The experimental data (black crosses) together with an exponential (magenta) and a power-law (blue) fit is shown with the large-lag standard error (grey). The green circles show the auto-correlation extracted from the simulated data.
	}\label{fig:angle-acf-log}\label{fig:model-angle-acf}
\end{figure}

\begin{figure}[h] 
	\begin{center}
	  \showimg{ \includegraphics[scale=0.26,angle=-90]{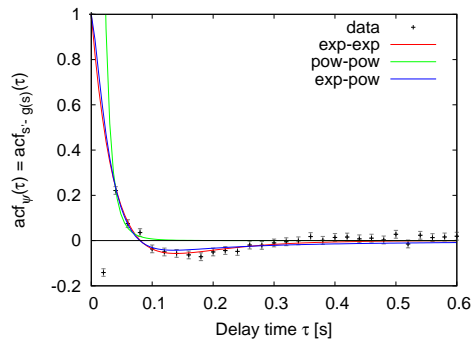} }
	\end{center}
	\caption{
	{\bf Auto-correlation of the non-deterministic speed changes $\psi(t)$.}
	The auto-correlation function of $\psi(t)=ds(t)/dt-g(s(t))$  estimated from the experimental data (dots) with two times the large-lag standard error (grey) and three fitted approximations: difference of 2 exponentials (red), difference of 2 power-laws (green),
	difference of exponential and power-law (blue).
	\added{The outlier at $\tau=\unit[0.02]{s}$ is a discretization artifact due to the finite resolution of the data (see \cite{Dieterich2008}).}
	}\label{fig:acc-acf}
\end{figure}

  For the two stochastic parts of the Langevin equations, we estimated the normalised auto-correlation functions from the data.
  The turning angle auto-correlation is approximated by a steep power-law as seen in Fig.~\ref{fig:angle-acf-log},
  which in this case is preferable to the alternative fit by a simple exponential decay.
  By subtraction of our approximation for the deterministic term $g(s)$ from the observed speed changes $ds/dt$ in
  Eq.~(\ref{eq:completemodel-s}) we estimated the distribution and auto-correlation of the noise term
  $\psi(t)=ds(t)/dt-g(s(t))$.
  In order not to overestimate the noise term, additive discretization errors of an approximate size
  of $\sigma_\mathrm{error}=\Delta x/{\Delta t}^2$ due to the finite
  resolution $\Delta x=\unit[10^{-3}]{m}$ of the cameras have been accounted for,
  giving the variance $\sigma_\psi^2=\sigma_{\psi^\mathrm{noisy}}^2-\sigma_\mathrm{error}^2$.
  The noise term $\psi(t)$ is well approximated by Gaussian noise with an auto-correlation function
  $\mathrm{acf}^{e-e}_\psi(\tau)= a e^{-\lambda_1 \tau} + (1-a) e^{-\lambda_2 \tau}$ (see Fig.~\ref{fig:acc-acf}).
  While an auto-correlation function of the shape of
  $\mathrm{acf}^{p-p}_\psi(\tau)= b (\tau+1)^{-p_1} + (1-b) (\tau+1)^{-p_2}$
  can be exluded, a difference between an exponential and a power-law
  $\mathrm{acf}^{e-p}_\psi(\tau)= c e^{-\lambda_3 \tau} + (1-c) (\tau+1)^{-p_2}$
  is not significantly worse than $\mathrm{acf}^{e-e}_\psi$.
  For our model we chose the simple difference of exponentials $\mathrm{acf}^{e-e}$.
 
  As the observed anti-correlation between delays of $\unit[0.1]{s}>\tau>\unit[0.3]{s}$ happens on a time scale
  which is too short to be an effect of the boundedness
  of the experiment or of residual effects of the presence of the foraging wall\cite{ourPRL},
  it is unclear where the anti-correlation comes from.
  One could speculate that it might be the result of a stabilising mechanism in the bumblebee dynamics.

\subsection*{Validation}

  Given all the parameters of the full model (see Materials and
  Methods) estimated by minimizing the mean squared errors,
  we used them to generate artificial bumblebee trajectories,
  as follows: We simulated the dynamics using an
  Euler-Maruyama scheme with noise terms $\xi_s(t)$,$\psi(t)$.
  In rare cases where the Gaussian noise $\psi(t)$ would lead to a negative speed despite the positive drift $g(s)$ for $s<s_0$, we enforce a non-negative speed by setting $s(t)=0$.
  We correlated the noise terms in advance by modifying their power
  spectral density in the following way:
  we take uncorrelated noise of the wanted distribution,
  multiply its Fourier transform with the root of the desired power spectral density corresponding to
  our approximate auto-correlation function and then transform
  back\cite{NumericalRecipes}.  To deal with the speed dependence of
  the turning angle noise $\xi_s(t)$ we first correlate Gaussian noise
  and afterwards scale with $\sigma_{\beta}(s)$ at each time step in the integration scheme.
  While this does not reproduce the auto-correlation of the turning angle exactly,
  the error made is less than the errors from the estimation of $\mathrm{acf}_{\beta}$.
  \begin{figure}[h] 
	\begin{center}
		\showimg{ \includegraphics[scale=0.55,angle=0]{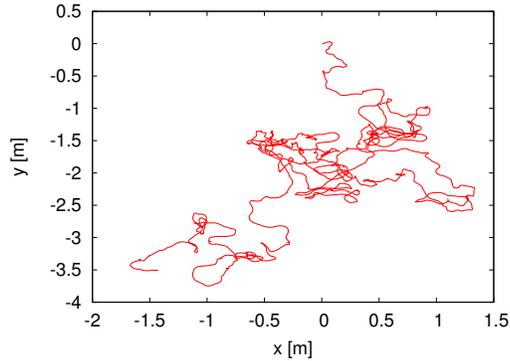} }
	\end{center}
	\caption{
	{\bf Simulated trajectory of a bumblebee.}
	 The complete model (Eqs.~(\ref{eq:completemodel-beta},\ref{eq:completemodel-s})) is simulated for \unit[200]{s} ($=10^5$ time steps)
	 with an Euler-Maruyama scheme using already correlated noise for $\xi$ and $\psi$.
	}\label{fig:model-traj}
\end{figure}
  A sample trajectory of a bumblebee simulated for \unit[200]{s} using $10^5$ time steps is shown in Fig.~\ref{fig:model-traj}.
  Using the generated data we checked the validity of the model by comparison to the experimental data of all bumblebees.

\begin{figure}[h] 
	\begin{center}
		\showimg{ \includegraphics[scale=0.55,angle=0]{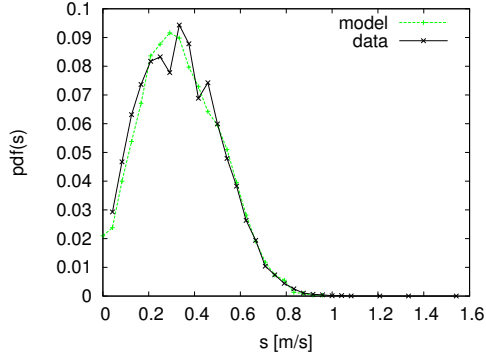} }
	\end{center}
	\caption{
	{\bf Comparison of the speed distributions.}
	The green (dashed) line shows the probability density $\mathrm{pdf}(s)$ extracted from the simulated data, the black (solid) line shows the experimental data of all bumblebees ($\approx45000$ data points).
	}\label{fig:model-speed-pdf}
\end{figure}

\begin{figure}[h] 
	\begin{center}
	 	\showimg{ \includegraphics[scale=0.55,angle=0]{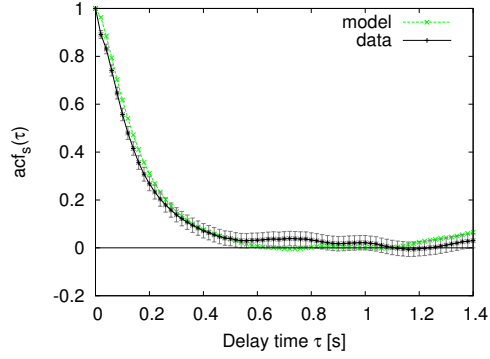} }
	\end{center}
	\caption{
	{\bf Auto-correlation of bumblebee speed.}
	The green (dashed) line shows the auto-correlation extracted from the simulated data, the black (solid) line from the experimental data with two times the large-lag standard error (grey).
	}\label{fig:model-speed-acf}
\end{figure} 

  Figure~\ref{fig:model-speed-pdf} compares the probability density function $\mathrm{pdf}(s)$
  of the speed extracted from the simulated data with the experimental data.
  The auto-correlation functions of the speed and turning angle are shown in Figures~\ref{fig:model-speed-acf}
  and \ref{fig:model-angle-acf}.
  Considering the number of rough approximations we have made for constructing our model,
  the agreement between simulation results and experimental data is very good.

\subsection*{Summary}

  We generalised a reorientation model which is often used to describe the correlated random walk of animals
  by explicitly modelling accelerations via Langevin equations.
  Analysing movement data from bumblebees, we extracted information on the deterministic and stochastic terms
  of Eqs.~(\ref{eq:generalmodel-beta},\ref{eq:generalmodel-s}).
  Simulations of our model and comparison to the data have shown that the resulting model agrees very  well
  with the experimental data despite the approximations we made for the model.
  With the estimation of the turning angle drift $h(\beta)$ we found that while the usual assumption
  of i.i.d.\ turning angles is not valid in our case, the lack of a non-trivial drift
  and the weak auto-correlation of $\xi_s$ are consistent with the usual reorientation model.
  However, our generalised model exhibits significant differences in the non-trivial deterministic part $g(s)$ of the speed
  change $ds/dt$ \changed{and} the speed dependence of the turning angles\deleted{ and the correlations in the noise terms}.
  In terms of active Brownian particle models\cite{Geier2012,Geier2011} we described the two-dimensional bumblebee movement by a particle
  with a non-linear friction term $g(s)$ depending and acting only on the speed, driven by multiplicative coloured noise with
  different correlations for the angle component and the speed component of the velocity.
  While this combination of complications might make it difficult to
  treat the system analytically,
progress into this direction has been
made\cite{Peruani2007,Lindner2010}.  \added{We remark that one could
ignore the fast decaying auto-correlations of $\xi_s$ and $\psi(t)$ if
one is not interested in the dynamics for short times, thus
simplifying the model by using uncorrelated noise terms, since the
effect of the noise autocorrelations on the long time dynamics is
negligible.}

Given that the experiment which yielded our data is rather small and
provided the bumblebees with an artificial environment,
  it would be interesting to apply our new model to free-flying bumblebees to reveal how much the results depend on the
  specific set-up. This would clarify whether the flight behaviour seen in the laboratory experiment survives as a
  flight mode for foraging in a patch of flowers in an intermittent model,
  with an additional flight mode for long flights between flower
  patches.  The analysis of data from other flying insects and birds
  by using our model could be interesting in order to examine
  whether the piecewise linear nature of the speed drift and the
  trivial drift of the turning angle are a common feature.
  In view of understanding the small-scale bio-mechanical origin of flight
  dynamics, our model might serve as a reference point
  for any more detailed dynamical modelling. That is, we would expect
  that any more microscopic model should reproduce our dynamics
  after a suitable coarse graining over relevant degrees of freedom.

\section*{Materials and Methods}
\subsection*{Experimental Data}

  In this experiment 30 bumblebees ({\it Bombus terrestris}) were trained to forage individually in a roughly cubical flight arena with an approximate side length of \unit[0.75]{m}.
  Figure~\ref{fig:cage} shows a diagram of the arena together with data from a typical flight path of a bumblebee.
  The flight arena included a $4\times4$ grid of artificial flowers on one of the walls.
  Each of the 16 flowers consisted of a landing platform, a yellow square floral marker and a replenishing food source where syrup was offered.
  For the analysis presented in this paper all data in zones ($\unit[7]{cm}\times\unit[9]{cm}\times\unit[9]{cm}$)
  around the flowers has been removed in order to analyse the search behaviour while foraging excluding the interaction with food sources.
  The 3D flight trajectories of the bumblebees were tracked by two cameras with a temporal resolution
  of $\Delta t=\unit[0.02]{s}$.
  Each bumblebee was approximated as a point mass with a spatial resolution of $\unit[0.1]{cm}$:
         its position was estimated by the arithmetic mean of all image pixels
         corresponding to the bumblebee via background subtraction.
  In total $\approx49000$ data points were used for the analysis.
  For individual bumblebees an average of $51$ search trajectories between flower zones
         have been sampled and analysed.
         The thorax widths of the bumblebees have a mean of $\unit[5.6]{mm}$
         and a standard deviation of $\unit[0.4]{mm}$.

  For calculating auto-correlations small gaps in the time series have been interpolated linearly.
  As the number of gaps was small the correlations for short times were not affected,
  however, the interpolation increased the usable data for long time delays.
  Trajectories were split at larger gaps, e.g., when entering a flower zone,
  to exclude correlations induced by flower visits. 
  
  For a discussion of the influence of the boundedness of the flight arena and for the analysis of the foraging dynamics under varying environmental conditions see \cite{ourPRL}.
  More details on the experimental setup can be found in \cite{Ings2008,Ings2009}.

\subsection*{Estimated Model Parameters}

  The full set of parameters estimated from the data set which was used for the simulation is given here.
  For the deterministic drift of the speed the change of slope is at $s_0=\unit[0.275]{m/s}$
  while the slopes are $d_1=0.16$ and $d_2=0.06$.
  The parameters for the standard deviation $\sigma_\xi(s)$ of the angle noise are
  $c_1=126^\circ$, $c_2=\unit[12]{s/m}$, $c_3=12.5^\circ$
  and its auto-correlation is given by $\mathrm{acf}_\beta(\tau)=(\tau+1)^{-1.5476}$.
  The non-deterministic changes $\psi(t)$ of the speed are assumed to be normally distributed
  with standard deviation $\sigma_\psi=\unit[3.52]{m/s^2}$ 
  and auto-correlated according to $\mathrm{acf}^{e-e}_\psi(\tau)$ where $a=1.44$, $\lambda_1=25.5$ and $\lambda_2=10.7$.

\subsection*{Speed Dependence of Turning Angles}
  A simple model showing a dependence of the turning angles on the speed (see Fig.~\ref{fig:schematic})
  is given in the following.
  Assume that the velocity of an animal changes at each time step $\Delta t$ by an acceleration vector
  which is given by a binormal i.i.d.\ random vector with variance $\sigma^2$ in both directions.
  The turning angle $\beta$ between $v_t$ and $v_{t+\Delta t}$ then depends on the quotient $\eta_t:=s_t/(\sqrt{2}\sigma)$
  between the former speed $s_t=|v_{t}|$ and the noise strength $\sigma$.
  By changing to the comoving frame of the animal and integrating out $s_{t+\Delta t}$
  the distribution $\rho(\beta)$ of the turning angle is given by:
  $$\rho(\beta)=\frac{e^{-\eta^2}}{2\pi}
              + \frac{e^{-\eta^2\sin^2(\beta)}}{2\sqrt{\pi}}\eta\cos(\beta)(1+\mathrm{erf}(\eta\cos(\beta)))$$
  for $-\pi\leq\beta\leq\pi$.
  With vanishing speed $s(t)=\eta(t)=0$ the first term gives a uniform distribution as expected,
  and for $\eta(t)\rightarrow\infty$ the distribution sharply peaks at $\beta=0$
  with its variance $\sigma_\beta$ approaching $0$, similar to the behaviour of the simpler von Mises distribution.
  As the experimental bumblebee data does not show a decay to $\sigma_\beta=0$ but to a finite value
  (see Fig.~\ref{fig:beta-s-loglog}), this simple model does not hold:
  therefore the accelerations have to be modelled as speed-dependent.

\section*{Acknowledgements}
  We thank Thomas C.~Ings and Lars Chittka for providing us with their experimental data
	and for their helpful	comments.

\bibliography{all.bib}

\end{document}